\documentstyle[prb,aps]{revtex}
\begin{document}
\input{psfig}
\twocolumn[\hsize\textwidth\columnwidth\hsize\csname @twocolumnfalse\endcsname

\title{Role of Secondary Motifs in Fast Folding Polymers: A Dynamical
Variational Principle}

\author{Amos Maritan$^1$, Cristian Micheletti$^1$ and Jayanth R. Banavar$^2$}
\vskip 0.3cm
\address{(1) International School for Advanced Studies (S.I.S.S.A.) - INFM,
Via Beirut 2-4, 34014 Trieste, Italy and \\
the Abdus Salam International Centre for Theoretical Physics.}
\address{(2) Department of Physics and Center for Materials Physics,
104 Davey Laboratory, The Pennsylvania State University, University
Park, Pennsylvania 16802}
\date{\today}
\maketitle
\begin{abstract}
A fascinating and open question challenging biochemistry, physics and
even geometry is the presence of highly regular motifs such as
$\alpha$-helices in the folded state of biopolymers and proteins.
Stimulating explanations ranging from chemical propensity to simple
geometrical reasoning have been invoked to rationalize the existence
of such secondary structures. We formulate a dynamical variational
principle for selection in conformation space based on the requirement
that the backbone of the native state of biologically viable polymers
be rapidly accessible from the denatured state. The variational
principle is shown to result in the emergence of helical order in
compact structures.
\end{abstract}
]

A fundamental problem in every day life is that of packing
with examples ranging from fruits in a grocery,
clothes and personal belongings in a suitcase,
atoms and colloidal particles in crystals and glasses, and
amino acids in the folded state of proteins. The
simplest problem in packing consists of determining the spatial
arrangement that accomodates the highest packing density of its
constituent entities with the result being a crystalline structure.
Besides packing considerations, dynamical effects
play a significant role when rapid packing/unpacking is entailed, as
in the formation of amorphous glasses where crystallization is
dynamically thwarted or in the more familiar suitcase problem.

Fast packing has been recognized as a central issue for biopolymers,
such as proteins, since the early work of Levinthal \cite{Levin}.
Further, the native conformations display extremely regular motifs,
such as $\alpha$-helices or $\beta$-sheets. In this Letter we
postulate a direct connection between the dynamics of rapid folding
and the emergence of secondary motifs in the native state
conformations. In fact, an intuitive approach to rapid and
reproducible folding might be to create neat patterns of lower
dimensional manifolds than the physical space and bend and curl them
into the final folded state. For proteins, secondary structures such
as $\alpha$-helices and $\beta$-sheets are indeed patterns in low
dimensions.

There are two key aspects distinguishing a protein from a generic
heteropolymer: the specially selected sequence of amino acids and the
three-dimensional structure that it folds reversibly into. For a given
target native structure, the selection mechanism in sequence space is
the principle of minimal frustration \cite{7}. The chosen sequences
are such that their target native states are reached through a
funnel-like landscape \cite{9} which facilitates the harmonious
fitting together of pieces to form the whole.

The three-dimensional structure impacts on the functionality of the
protein and a fascinating issue is the elucidation of the selection
mechanism in conformation space that picks out certain viable
structures from the innumerable ones with a given compactness. Earlier
studies have shown that there is a direct link between viable native
conformations and high designability \cite{LG}. Recently \cite{19}, it
was observed that the natural folds of proteins have a much larger
density of nearby structures than generic (artificial) conformations
of the same character and that the exceedingly large geometrical
accessibility of natural proteins may be related to the presence of
secondary motifs.

The realization that proteins have secondary structures arose with
early crystallographic studies and the brilliant deduction of Pauling
et al. \cite{2} of the ability of an $\alpha$-helix of the correct
pitch to accomodate hydrogen bonds, thus promoting its stability.
Inspired by the findings of Pauling, helix-coil transition models have
been used to study the thermodynamics of helix formation
\cite{hc}. The models encompass features that ensure the helical
nature of the low-energy states by assuming first that that monomers
can be in a helical state and by then introducing co-operative
interactions that favor helical regions. It is interesting to note,
however, that the number of hydrogen bonds is nearly the same when a
sequence is in an unfolded structure in the presence of a polar
solvent or in its native state rich in secondary structure content
\cite{4}. It has also been suggested that the $\alpha$-helix is an
energetically favorable conformation for main-chain atoms but the
side-chain suffers from a loss of entropy \cite{4,18}. Nelson {\em et
al.} \cite{21} have shown both numerically and experimentally that
non-biological oligomers fold reversibly like proteins into a specific
three-dimensional structure with high helical content driven only by
solvophobic interactions. Recent studies have attempted to explain
the emergence of secondary structure from geometrical principles
rather than invoking detailed chemistry. Despite the concerted efforts
of several groups, a simple general explanation remains elusive. In
particular, the work of Yee {\em et al.} \cite{3}, Hunt {\em et al.}
\cite{4}, and Socci {\em et al.} \cite{5} have shown that compactness
alone can only account for a small secondary structure content. These
facts are also corroborated by the recent study of the kinetics of
homopolymer collapse, where no evidence was found for the formation of
local regular structures \cite{halper}.

We propose a selection mechanism in structure space in the form of a
variational principle postulating that, {\em among all possible native
conformations, a protein backbone will attain only those which are
optimal under the action of evolutionary pressure favouring rapid
folding}. Our goal is to elucidate the role played by the bare native
backbone independent of the selection in sequence space and hence of
the (imperfectly-known) inter-amino-acid potentials. We therefore
choose to employ a Go-like model \cite{12} with no other interaction
that promotes or disfavours secondary structures. The model is a
sequence-independent limiting case of minimal frustration \cite{7}
which, for a given target native state conformation, favours the
formation of native contacts -- the energy of a sequence in a
conformation is simply obtained as the negative of the number of
contacts in common with the target conformation. We will consider two
non-consecutive amino acids to be in contact if their separation is
below a cutoff $r_0 = 6.5$ \AA (the results are qualitatively similar
when slightly different values of $r_0$ in the range $6-8$ \AA are
chosen).

The energy of structure $\Gamma$ in the Go model is given by
\begin{equation}
H(\Gamma) = - {1 \over 2} \sum_{i, j} \Delta_{i,j}(\Gamma)
\Delta_{i,j}(\Gamma_0)
\label{eqn:ham}
\end{equation}

\noindent where the sum is taken over all pairs of amino acids,
$\Gamma_0$ is the target structure, $\Delta_{i,j}(\Gamma)$ is the
contact map of structure $\Gamma$:
\begin{equation}
\Delta_{ij}(\Gamma) = \{
\begin{array}{l l}
1 &\ \ \ { R_{ij}<r_0 \mbox{ and } |i-j| >2;}\\
0 & \ \ \ \mbox{ otherwise, }
\end{array}
\end{equation}

\noindent where $R_{ij}$ is the distance of amino acids $i$ and $j$.

 The polypeptide chain is modelled as a chain of beads subject to
steric constraints \cite{17,19}. We adopted a discrete
representation similar to the one of Covell and Jernigan \cite{17},
in which each bead occupies a site of an FCC lattice with lattice
spacing equal to 3.8 \AA. Such a representation is able to describe
the backbone of natural proteins to better that 1 \AA \ rmsd per
residue (equal to the best experimental resolution) and preserves typical
torsional angles. All discretized structures were subject to a
suitable constraint: any two non-consecutive residues cannot be closer
than $4.65$ \AA\ due to excluded volume effects and the distance
between consecutive residues can fluctuate between 2.6 \AA $< d <$ 4.7
\AA. Such constraints were determined by an analysis of the
coarse-grainings of several proteins of intermediate length ($\approx
100$ residues). In order to enforce a realistic global compactness
for a backbone of length $L$, the number of contacts in all the target
structures considered was chosen \cite{cont} to be around $N=1.9L$
while, locally, no residue was allowed to make contact with four or
more consecutive residues.

In order to assess the validity of the variational principle, it is
necessary to evaluate the typical time, $t(\Gamma_0)$, taken to fold
into a given target structure, $\Gamma_0$, followed by a selection of
the structures $\Gamma_0$, that have the smallest folding times.  To
do this, an initial set of ten conformations was generated by
collapsing a loose chain starting from random initial conditions. In
each case, we modified the random initial conformation by using Monte
Carlo dynamics: we move up to 3 consecutive beads to unoccupied
discrete positions that do not violate any of the physical constraints
and accept the moves according to the standard Metropolis rule. The
energy is given by eq. (\ref{eqn:ham}), while the temperature for the
MC dynamics was set to 0.35. This value was chosen in preliminary runs
so that it was higher than the temperature \cite{7} below which the
sequence is trapped in metastable states but comparable to the folding
transition temperature so that conformations with significant overlap
with the native state are sampled in thermal equilibrium.

For each structure, as a measure of the folding time we took the
median over various attempts (typically 41) of the total number of
Monte Carlo moves necessary to form a pre-assigned fraction of native
contacts, typically 66\%, starting from a random conformation. Our
results were unaltered on increasing this fraction to 75\%; indeed,
this fraction could be progressively increased towards 100\% with
successive generations without increase in the computational cost
since better and better folders are obtained.

A new generation of ten structures is created by ``hybridizing'' pairs
of structures of the previous generation ensuring that structures with
small folding times are hybridized more and more frequently as the
number of generations, $g$, increases \cite{14}. To do this, each of
the two distinct parent structures to be hybridized, $\Gamma_1$ and
$\Gamma_2$ are chosen with probability proportional to
$\exp[-(g-1)*ft)/1000]$, where $g$ is the index of the current
generation (initially equal to 1), $ft$ is the median folding
time. Then, a hybrid map is created by taking the union of the two
parent maps:
\begin{equation}
\Delta_{ij}^{Union} = \max(\Delta_{ij}(\Gamma_1),\Delta_{ij}(\Gamma_2))\ .
\end{equation}

\noindent Because it is not guaranteed that $\Delta^{Union}$
corresponds to a three-dimensional structure obeying the same physical
constraints as $\Gamma_1$ and $\Gamma_2$, the corresponding hybrid
$\Gamma$ is constructed by taking one of the two parent structures (or
alternatively a random one) as the starting conformation and carrying
out MC dynamics favouring the formation of each of the contacts in the
union map (i.e. using eq. (1) with $\Delta_{ij}(\Gamma_0)$ substituted
by $\Delta_{ij}^{Union}$). The dynamics is carried out starting from a
temperature of $0.7$ and then decreasing it gradually over a
sufficiently long time (typically thousands of MC steps) to achieve
the maximum possible overlap with the union map, while simultaneously
maintaining the realistic compactness. The resulting hybrid structure
is typically midway between the two parent structures, in that it
inherits native contacts from both of them. We adopted the following
definition in order to obtain an objective and unbiased way to
quantitatively estimate the presence of secondary content: a given
residue, $i$ was defined to belong to a secondary motif if, for some
$j$, one of these conditions held:

\begin{eqnarray}
a)\ \Delta_{i-1,j-1}&=&\Delta_{i,j}=\Delta_{i+1,j+1}=\Delta_{i,j+1}\nonumber \\
        &=& \Delta_{i+1,j+2}=\Delta_{i-1,j}=1;\nonumber \\
b)\ \Delta_{i+1,j-1}&=&\Delta_{i,j}=\Delta_{i-1,j+1}=\Delta_{i,j+1}\nonumber \\
        &=& \Delta_{i+1,j}=\Delta_{i-1,j+2}=1.\nonumber
\end{eqnarray}

\noindent The former [latter] identifies the presence of helices and
parallel [anti-parallel] $\beta$ sheets in natural proteins, which can
be identified by the visual inspection of contact matrices and appears
as thick bands parallel or orthogonal to the diagonal.

\noindent The upper plot of Fig. 1 shows the decrease of the typical
folding time over the generations for chains of length 25, while the
middle panel shows the accompanying increase in the number of residues
in secondary motifs (secondary content). The bottom panel shows a
milder decrease of the contact order (i.e. a larger number of
short-range contacts) as the generations evolved, in agreement with
the experimental findings of Plaxco {\em et al.}\cite{15}

One of the optimal structures of length 25 is shown in Figure
\ref{fig:fig2}a. Due to the absence of any chirality bias in our
structure space exploration, the helix does not have a constant
handedness. The signature of the secondary motifs in the optimal
structures is clearly visible in the contact maps of Figure
\ref{fig:fig3}, which are not sensitive to structure chirality.
Strikingly, the variational principle selects conformations with
significant secondary content as those facilitating the fastest
folding. The correlation of the emergence of secondary structures with
decrease of folding times is shown in the plot of
Fig. \ref{fig:trend}. We verified that the hybridization procedure is
not biased towards low contact order by iterating it for various
generations and hybridizing the structures at random. Even after
dozens of generations, the generated structures had secondary contents
of about 1/3-1/4 of the true extremal structures.

The very high secondary content in optimal conformations was found to
be robust against changes in chain length or compactness of the target
structure. On requiring that the structure be more compact, bundles of
helices emerge [see Fig. \ref{fig:fig2}b] along with an increase in
contact order, signalling the presence of some longer range contacts,
which are necessitated in order to accomodate the shorter radius of
gyration. It is noteworthy that our calculations lead predominantly
to $\alpha$-helices and not $\beta$ sheets, a fact accounted for by
the demonstration that steric overlaps and the associated loss of entropy
lead to the destabilization of helices in favor of sheets \cite{18},
the appearance of such sheets only in sufficiently long proteins
\cite{22} and the much slower folding rate of $\beta$-sheets
compared to $\alpha$-helices \cite{slow}. It is remarkable that the same
requirement of rapid folding is sufficient to lead to a selection in
both sequence and structure space underscoring the harmony in the
evolutionary design of proteins. The results and strategies presented
here ought to be applicable in protein-engineering contexts, for
example by ensuring optimal dynamical accessibility of the backbone of
proteins. A systematic collection of the rapidly-accessible
structures of various length should also lead to the creation of
unbiased libraries of protein folds.

{\bf Acknowledgements} This work was supported by INFM, INFN sez. di
Trieste, NASA, NATO and The Donors of the Petroleum Research Fund
administered by the American Chemical Society. We thank F. Seno and A.
Trovato for useful discussions.\\

\begin{figure}
\label{fig:fig1}
\centerline{\psfig{figure=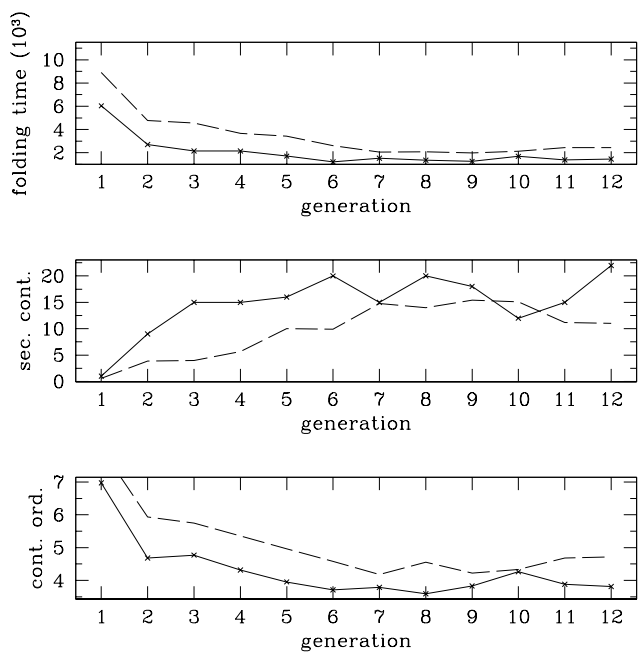,width=3.5in}}
\caption{Evolution of the median folding time (measured in Monte Carlo
steps), secondary structure content and contact order as a function of
the number of generations in the optimization      algorithm for compact
structures of length $L=25$. The dashed curve denotes an average over
all ten structures in a given generation, whereas the solid curve
shows the behaviour of the structure at each generation with the
fastest median folding time. Analogous results are obtained for other
runs and for other values of $L$. The dramatic decrease of folding
time is accompanied by an equally significant increase in the
secondary content.}
\end{figure}

\begin{figure}
\centerline{\psfig{figure=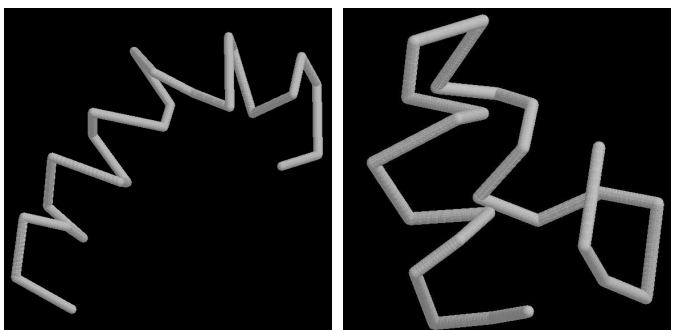,width=3.5in}}
\caption{ a) RASMOL plot of a structure with very low median folding
time and $L=25$. b) Structure with very low median folding time,
$L=25$ and higher compactness (all target conformations were
constrained to have a radius of gyration smaller than 6.5
\AA). Optimal compact structures correspond to helices packed
together, as observed in naturally occurring proteins.}
\label{fig:fig2}
\end{figure}

\begin{figure}
\centerline{\psfig{figure=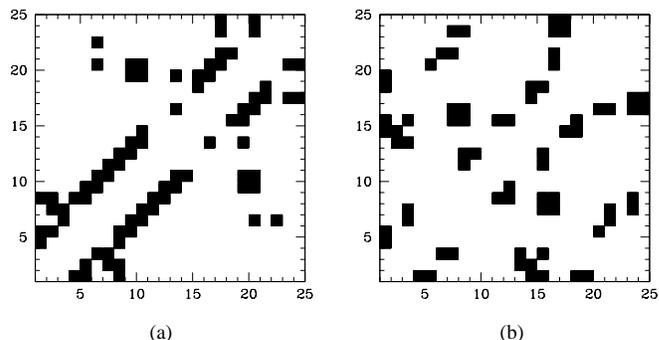,width=3.5in}}
\caption{The panel on the left [right] shows the contact map of a
structure with a very low [average] median folding time. The signature
of helices in map (a) is shown by the thick bands parallel to the
diagonal, while no such patterns are observed in the matrix (b).}
\label{fig:fig3}
\end{figure}

\begin{figure}
\centerline{\psfig{figure=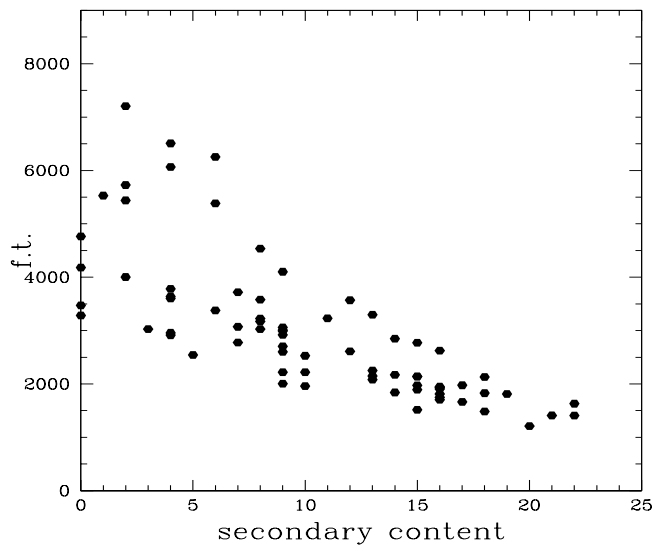,width=3.5in,height=3.0in}}
\caption{Scatter plot of folding time versus secondary content for
structures of length 25 collected over several generation of the
optimization algorithm.}
\label{fig:trend}
\end{figure}


\begin{references}

\bibitem{Levin} C. Levinthal, J. Chim. Phys. {\bf 65}, 44 (1968).

\bibitem{7} J. D. Bryngelson and P. G. Wolynes, {\em Proc. Natl.
Acad. Sci. USA} {\bf 84}, 7524-7528 (1987); J. D. Bryngelson,
J. N. Onuchic, J. N. Socci and P. G. Wolynes, {\em Proteins:
Struc. Funct. Genet.} {\bf 21}, 167-195 (1995).

\bibitem{9} P. E. Leopold, M. Montal and J. N. Onuchic, {\em
Proc. Natl. Acad. Sci. USA} {\bf 89}, 8721-8725 (1992); P. G. Wolynes
J. N. Onuchic and D. Thirumalai, {\em Science} {\bf 267}, 1619-1620
(1995); J. N. Onuchic, Z. Luthey Schulten and P. G. Wolynes, {\em
Ann. Rev. Phys. Chem.} {\bf 48}, 545-600 (1997); K. A. Dill and
H. S. Chan, {\em Nature Structural Biology} {\bf 4}, 10-19 (1997).

\bibitem{LG} H. Li, R. Helling, C. Tang and N. Wingreen, {\em Science}
{\bf 273}, 666-669 (1996); N. E. G. Buchler and R. A. Goldstein, {\em
Proteins: Struc. Funct. Genet.} {\bf 34}, 113-124 (1999);
C. Micheletti, A. Maritan, J. R. Banavar and F. Seno, {\em
Phys. Rev. Lett.} {\bf 80}, 5683 (1998); C. Micheletti, A. Maritan and
J. R. Banavar, {\em J. Chem. Phys.} {\bf 110}, 9730 (1999).



\bibitem{19} C. Micheletti, J. R. Banavar, A. Maritan and F. Seno,{\em
Phys. Rev. Lett.} {\bf 82}, 3372-3375 (1999).

\bibitem{2} L. Pauling, R. B. Corey and H. R. Branson,
{\it Proc. Nat. Acad. Sci.} {\bf 37}, 205-208 (1951).

\bibitem{hc} B. H. Zimm and J. Bragg, {\em J. Chem. Phys.}, {\bf 31},
526 (1959); O. B. Ptitsyn and A. M. Skvortsov, {\em Biophys.} {\bf
10}, 1007 (1965); I. M. Lifshitz, A. Y. Grosberg and A. R. Khokhlov, {\em
Rev. Mod. Phys.}, {\bf 50}, 683 (1978).


\bibitem{4} N. G. Hunt, L. M. Gregoret and F. E. Cohen,
{\it J. Mol. Biol.} {\bf 241}, 214-225 (1994).

\bibitem{18} R. Aurora, T. P. Creamer, R. Srinivasan and G. D. Rose,
{\em J. Mol. Biol.} {\bf 272}, 1413-1416 (1997).

\bibitem{21} J. C. Nelson, J. G. Saven, J. S. Moore, and P. G. Wolynes,
{\em Science} {\bf 277}, 1793-1796 (1997).

\bibitem{3} D. P. Yee, H. S. Chan, T. F. Havel and K. A. Dill,
{\it J. Mol. Biol.} {\bf 241}, 557-573 (1994).

\bibitem{5} N. D. Socci, W. S. Bialek, and J. N. Onuchic,
{\it Phys. Rev. E} {\bf 49}, 3440-3443 (1994).

\bibitem{halper} A. Halperin and P. M. Goldbart, Phys. Rev. E, in
press (cond-mat/9905306).

\bibitem{12} N. Go, {\em Macromolecules} {\bf 9}, 535-541 (1976).


\bibitem{17} D. G. Covell and R. Jernigan {\it Biochem.} {\bf 29},
3287 (1990).

\bibitem{cont} The native state structures of monomeric
proteins of length between 50 and 200 show an excellent correlation of
this form, when two non-consecutive amino acids along the sequence are
defined to be in contact when they are within 6.5 \AA\ of each
other.

\bibitem{14} J. H. Holland, {\em Adaptation in natural and artificial
systems}, MIT press ed. (1992).

\bibitem{15} K. M. Plaxco, K. T. Simons and D. Baker, {\it
J. Mol. Biol.} {\bf 277}, 985-994 (1998).

\bibitem{22} A. P. Capaldi and S. E. Radford, S. E., {\em
Curr. Op. Str. Biol.} {\bf 8}, 86-92 (1998).

\bibitem{slow} V. Mu\~noz, E. R. Henry, J. Hofrichter and W. A. Eaton,
{\em Proc. Natl. Acad. Sci. USA}, {\bf 95}, 5872 (1998).
\end{references}
\end{document}